\documentclass[structabstract]{aa}

\usepackage{graphicx}
\usepackage{natbib}

\newcommand{\Msun}{\ensuremath{\,{M}_\odot}}                      
\newcommand{\Rsun}{\ensuremath{\,{R}_\odot}}                      
\newcommand{\Teff}{\ensuremath{T_{\rm eff}}}                      
\newcommand{\Mjup}{\ensuremath{\,{M}_{\rm Jup}}}                  
\newcommand{\Rjup}{\ensuremath{\,{R}_{\rm Jup}}}                  
\newcommand{\Teq}{\ensuremath{T_{\rm eq}^{\,\prime}}}             
\newcommand{\safronov}{\ensuremath{\Theta}}                       
\newcommand{\kms}{\,km\,s$^{-1}$}                                 
\newcommand{\mss}{\,m\,s$^{-2}$}                                  
\newcommand{\as}{\ensuremath{^{\prime\prime}}}                    
\newcommand{\am}{\ensuremath{^\prime}}                            
\newcommand{\pjup}{\ensuremath{\,\rho_{\rm Jup}}}                 
\newcommand{\psun}{\ensuremath{\,\rho_\odot}}                     

\newcommand{\mcc}[1]{\multicolumn{3}{c}{#1}}
\newcommand{\er}[3]{\ensuremath{#1^{+#2}_{-#3}}}

\newcommand{\ermcc}[5]{\mcc{\ensuremath{{#1\,^{+#2}_{-#3}}\,^{+#4}_{-#5}}}}

\usepackage{amsmath,amssymb,graphicx}
\begin{document}
\title{Physical properties of the WASP-67 planetary system from multi-colour photometry\thanks{Based on data collected with GROND
at the MPG 2.2-m telescope and DFOSC at the Danish 1.54-m
telescope.}}
\titlerunning{Physical properties of WASP-67\,b}

   \author{
          L. Mancini\inst{1} 
          \and
          J. Southworth\inst{2}
          \and
          S. Ciceri\inst{1}
          \and
          S. Calchi Novati\inst{3,4}
          \and
          M. Dominik\inst{5}
          \and
          Th. Henning\inst{1}
          \and
          U.~G. J{\o}rgensen\inst{6,7}
          \and
          H. Korhonen\inst{8,6,7}
          \and
          N. Nikolov\inst{9}
          \and
          K.~A. Alsubai\inst{10}
          \and
          V. Bozza\inst{4,11}
          \and
          D.~M. Bramich\inst{12}
          \and
          G. D'Ago\inst{4,11}
          \and
          R. Figuera Jaimes\inst{5,13}
          \and
          P. Galianni\inst{5}
          \and
          S.-H. Gu\inst{14,15}
          \and
          K. Harps{\o}e\inst{6,7}
          \and
          T.~C. Hinse\inst{16}
          \and
          M. Hundertmark\inst{5}
          \and
          D. Juncher\inst{6,7}
          \and
          N. Kains\inst{17}
          \and
          A. Popovas\inst{6,7}
          \and
          M. Rabus\inst{18,1}
          \and
          S. Rahvar\inst{19}
          \and
          J. Skottfelt\inst{6,7}
          \and
          C. Snodgrass\inst{20}
          \and
          R. Street\inst{21}
          \and
          J. Surdej\inst{22}
          \and
          Y. Tsapras\inst{21,23}
          \and
          C. Vilela\inst{2}
          \and
          X.-B. Wang\inst{14,15}
          \and
          O. Wertz\inst{22}
          }
       \institute{
   Max Planck Institute for Astronomy, K\"{o}nigstuhl 17, 69117 -- Heidelberg, Germany \\
             \email{mancini@mpia.de}
         \and
   Astrophysics Group, Keele University, Staffordshire, ST5 5BG, UK
         \and
   International Institute for Advanced Scientific Studies (IIASS), 84019 Vietri Sul Mare (SA), Italy
         \and
   Department of Physics, University of Salerno, Via Giovanni Paolo II, 84084 Fisciano, Italy
         \and
   SUPA, University of St Andrews, School of Physics \& Astronomy, North Haugh, St Andrews, KY16 9SS, UK
         \and
   Niels Bohr Institute, University of Copenhagen, Juliane Maries vej 30, 2100 Copenhagen \O, Denmark
         \and
   Centre for Star and Planet Formation, Geological Museum, {\O}ster Voldgade 5-7, 1350 Copenhagen, Denmark
         \and
   Finnish Centre for Astronomy with ESO (FINCA), University of Turku, V{\"a}is{\"a}l{\"a}ntie 20, FI-21500 Piikki{\"o}, Finland
         \and
   Astrophysics Group, University of Exeter, Stocker Road, EX4 4QL, Exeter, UK
         \and
   Qatar Foundation, PO Box 5825, Doha, Qatar
         \and
   Istituto Nazionale di Fisica Nucleare (INFN), Sezione di Napoli, Napoli, Italy
         \and
   Qatar Environment and Energy Research Institute, Qatar Foundation, Tornado Tower, Floor 19, P.O. Box 5825, Doha, Qatar
         \and
   European Southern Observatory, Karl-Schwarzschild-Stra{\ss}e 2, 85748 Garching bei M\"unchen, Germany
        \and
   Yunnan Observatories, Chinese Academy of Sciences, Kunming 650011, China
         \and
   Key Laboratory for the Structure and Evolution of Celestial Objects, Chinese Academy of Sciences, Kunming 650011, China
         \and
   Korea Astronomy and Space Science Institute, Daejeon 305-348, Republic of Korea
         \and
   Space Telescope Science Institute, 3700 San Martin Drive, Baltimore, MD 21218, USA
         \and
   Instituto de Astrof\'isica, Facultad de F\'isica, Pontificia Universidad Cat\'olica de Chile, Av. Vicu\~na Mackenna 4860,
   7820436 Macul, Santiago, Chile
         \and
   Department of Physics, Sharif University of Technology, P.\,O.\,Box 11155-9161 Tehran, Iran
         \and
   Max Planck Institute for Solar System Research, Justus-von-Liebig-Weg 3, 37077 G\"{o}ttingen, Germany
         \and
   Las Cumbres Observatory Global Telescope Network, 6740B Cortona Drive, Goleta, CA 93117, USA
         \and
   Institut d'Astrophysique et de G\'eophysique, Universit\'e de Li\`ege, 4000 Li\`ege, Belgium
         \and
   School of Physics and Astronomy, Queen Mary University of London, Mile End Road, London, E1 4NS, UK
         }


  \abstract
{The extrasolar planet WASP-67\,b is the first hot Jupiter
definitively known to undergo only partial eclipses. The lack of
the second and third contact point in this planetary system makes
it difficult to obtain accurate measurements of its physical
parameters.}
{By using new high-precision photometric data, we confirm that
WASP-67\,b shows grazing eclipses and compute accurate estimates
of the physical properties of the planet and its parent star.}
{We present high-quality, multi-colour, broad-band photometric
observations comprising five light curves covering two transit
events, obtained using two medium-class telescopes and the
telescope-defocussing technique. One transit was observed through
a Bessel-$R$ filter and the other simultaneously through filters
similar to Sloan $g^{\prime}r^{\prime}i^{\prime}z^{\prime}$. We
modelled these data using {{\sc jktebop}}. The physical parameters
of the system were obtained from the analysis of these light
curves and from published spectroscopic measurements.}
{All five of our light curves satisfy the criterion for being
grazing eclipses. We revise the physical parameters of the whole
WASP-67 system and, in particular, significantly improve the
measurements of the planet's radius ($R_{\mathrm{b}}=1.091 \pm
0.046 \, R_{\mathrm{Jup}}$) and density ($\rho_{\mathrm{b}}=0.292
\pm 0.036 \, \rho_{\mathrm{Jup}}$), as compared to the values in
the discovery paper ($R_{\mathrm{b}}=1.4_{-0.2}^{+0.3} \,
R_{\mathrm{Jup}}$ and $\rho_{\mathrm{b}}=0.16 \pm 0.08 \,
\rho_{\mathrm{Jup}}$). The transit ephemeris was also
substantially refined. We investigated the variation of the
planet's radius as a function of the wavelength, using the
simultaneous multi-band data, finding that our measurements are
consistent with a flat spectrum to within the experimental
uncertainties.}
{}


\maketitle

\section{Introduction}
\label{sec:1}
WASP-67\,b \citep{hellier2012} is a transiting extrasolar planet
(TEP), discovered by the SuperWASP group \citep{pollacco2006},
orbiting a K0\,V star ($V=12.5$\,mag) every $4.61$\,d. It is an
inflated ($\rho_{\mathrm{b}} \ll \rho_{\mathrm{Jup}}$) hot Jupiter
($a\sim0.05$\,AU) on a grazing orbit (impact parameter $b>0.9$),
causing the transit light curve to have an atypical V shape.
\citet{hellier2012} found that WASP-67\,b satisfies by $3\sigma$
the grazing criterion ($X=b+R_{\mathrm{b}}/R_{\star}>1$), which
makes it the first TEP definitively known to have a grazing
eclipse\footnote{Other TEPs which might undergo grazing eclipses
are WASP-34 \citep{smalley2011} and HAT-P-27/WASP-40
\citep{beky2011,anderson2011}.}. In this particular configuration,
the second and third contact points (e.g.\ \citealp{winn2010}) are
missing and the light curve solution becomes degenerate. This fact
hampers accurate measurements of the photometric parameters of the
system. Consequently, \citet{hellier2012} measured the radius of
the planet with a large uncertainty of $\sim 20\%$. In such cases,
high-quality light curves are mandatory in order to reduce the
error bars to levels similar to those of other known TEPs.


Here we present the first photometric follow-up study of WASP-67
since its discovery paper. The main aim of this study is to refine
the physical parameters of the system and ephemeris, setting the
stage for a more detailed study in the near future. WASP-67 is
located in field \#7 of the K2 phase of the NASA's \emph{Kepler}
mission\footnote{http://keplerscience.arc.nasa.gov/K2/}, and will
be observed continuously for approximately 80\,d in late 2015.

\section{Observations and data reduction}
\label{sec:2}

\begin{figure*}
\centering
\includegraphics[width=16cm]{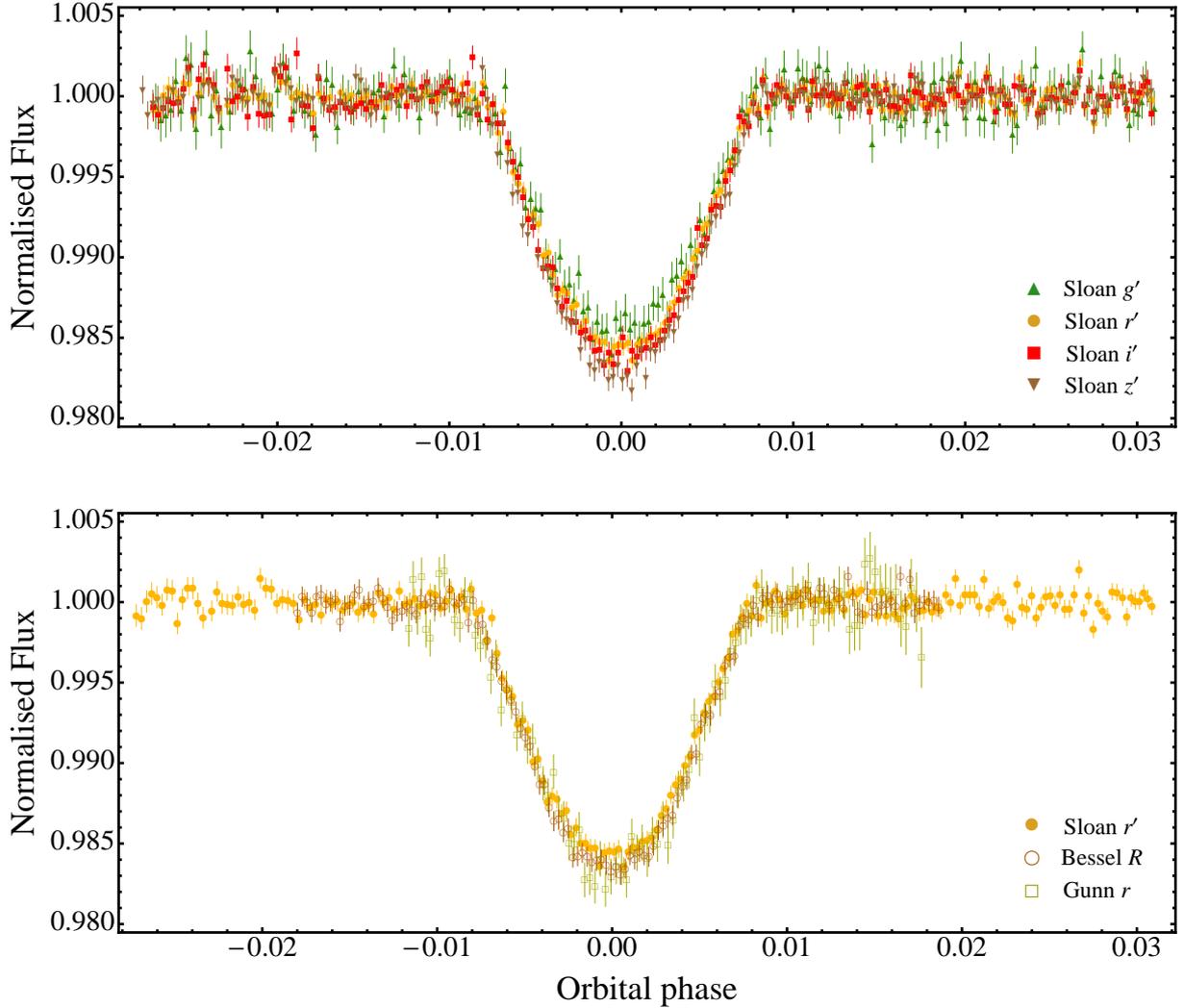}
\caption{Light curves of WASP-67\,b eclipses. \emph{Top panel}:
light curves obtained with GROND in
$g^{\prime}r^{\prime}i^{\prime}z^{\prime}$, showing how the
transit light curve shape changes with wavelength. The transit in
the $g^{\prime}$ band is shallower than the other bands as
expected for a grazing eclipse, as limb darkening is stronger at
bluer wavelengths. \emph{Bottom panel}: light curves obtained with
DFOSC in the $R$-band (June 2013, brown open circles), with GROND
in the $r^{\prime}$-band (June 2012, yellow points) and with the
Euler 1.2-m telescope in the $r$-band (July 2011, green open
squares, \citealt{hellier2012}). The light curves are superimposed
to highlight variations in transit shape between the three
measurements.} \label{Fig:01}
\end{figure*}

\begin{table*}
\caption{Details of the transit observations presented in this work.} %
\label{tab:01} %
\centering     %
\tiny          %
\setlength{\tabcolsep}{5pt}
\begin{tabular}{lcccrrcccccc}
\hline\hline
Instrument & Date of   & Start time & End time & $N_{\rm obs}$ & $T_{\rm exp}$ & $T_{\rm obs}$ & Filter & Airmass & Moon   & Aperture   & Scatter \\
          & first obs. & (UT)       & (UT)     &                    & (s)           & (s)           &        &    & illum. & radii (px) & (mmag) \\
\hline
GROND & 2012 06 04 & 03:00 & 10:50 & 162 & 70/90 & 110/120 & Sloan $g^{\prime}$ & $2.14 \rightarrow 1.01 \rightarrow 1.22$ & $ 98\%$ & 34, 50, 80 & 1.08 \\
GROND & 2012 06 04 & 03:00 & 10:50 & 162 & 70/90 & 110/120 & Sloan $r^{\prime}$ & $2.14 \rightarrow 1.01 \rightarrow 1.22$ & $ 98\%$ & 38, 60, 85 & 0.56 \\
GROND & 2012 06 04 & 03:00 & 10:50 & 162 & 70/90 & 110/120 & Sloan $i^{\prime}$ & $2.14 \rightarrow 1.01 \rightarrow 1.22$ & $ 98\%$ & 40, 60, 85 & 0.72 \\
GROND & 2012 06 04 & 03:00 & 10:02 & 230 & 70/90 & 110/120 & Sloan $z^{\prime}$ & $2.14 \rightarrow 1.01 \rightarrow 1.22$ & $ 98\%$ & 40, 60, 85 & 0.64 \\[2pt]
DFOSC & 2013 06 22 & 04:30 & 08:33 & 136 & 100 & 110 & Bessel $R$  & $1.12 \rightarrow 1.01 \rightarrow 1.17$ & $ 97\%$ & 20, 35,  55 & 0.48 \\[2pt]
\hline
\end{tabular}
\tablefoot{$N_{\rm obs}$ is the number of observations, $T_{\rm
exp}$ is the exposure time, $T_{\rm obs}$ is the observational
cadence, and `Moon illum.' is the fractional illumination of the
Moon at the midpoint of the transit. The aperture sizes are the
radii of the software apertures for the star, inner sky and outer
sky, respectively. Scatter is the r.m.s.a scatter of the data
versus a fitted model.}
\end{table*}

\begin{table}
\caption{Excerpts of the light curves of WASP-67: this table will
be made available at the CDS. A portion is shown here for guidance
regarding its form and content.}
\label{tab:02} %
\centering     %
\tiny          %
\begin{tabular}{lccrc}
\hline\hline
Telescope    & Filter & BJD (TDB) & Diff. mag. & Uncertainty  \\
\hline
ESO 2.2-m  & $g^{\prime}$ & 2456082.655745 &  0.00061 & 0.00043 \\
ESO 2.2-m  & $g^{\prime}$ & 2456082.657102 &  0.00142 & 0.00043 \\[2pt]
ESO 2.2-m  & $r^{\prime}$ & 2456082.655745 &  0.00083 & 0.00038 \\
ESO 2.2-m  & $r^{\prime}$ & 2456082.657102 &  0.00101 & 0.00033 \\[2pt]
ESO 2.2-m  & $i^{\prime}$ & 2456082.655745 &  0.00069 & 0.00041 \\
ESO 2.2-m  & $i^{\prime}$ & 2456082.657102 &  0.00117 & 0.00043 \\[2pt]
ESO 2.2-m  & $z^{\prime}$ & 2456082.653032 & -0.00041 & 0.00048 \\
ESO 2.2-m  & $z^{\prime}$ & 2456082.654390 & -0.00117 & 0.00048 \\[2pt]
DK 1.54-m  & $R$          & 2456465.694278 &  0.00066 & 0.00141 \\
DK 1.54-m  & $R$          & 2456465.695528 &  0.00033 & 0.00141 \\[2pt]
\hline
\end{tabular}
\end{table}

A complete transit of WASP-67\,b was observed on 2012 June 4
(Table\,\ref{tab:01}) using the \textbf{G}amma \textbf{R}ay Burst
\textbf{O}ptical and \textbf{N}ear-Infrared \textbf{D}etector
(GROND) instrument mounted on the MPG\footnote{Max Planck
Gesellschaft.} 2.2-m telescope, located at the ESO observatory in
La Silla (Chile). GROND is an imaging system capable of
simultaneous photometric observations in four optical (similar to
Sloan $g^{\prime}$, $r^{\prime}$, $i^{\prime}$, $z^{\prime}$) and
three NIR ($J,\, H,\, K$) passbands \citep{greiner2008}. Each of
the four optical channels is equipped with a back-illuminated
$2048 \times 2048$ E2V CCD, with a field of view of $5.4^{\prime}
\times 5.4^{\prime}$ at $0.158^{\prime\prime}\rm{pixel}^{-1}$. The
three NIR channels use $1024 \times 1024$ Rockwell HAWAII-1 arrays
with a field of view of $10^{\prime}\times 10^{\prime}$ at
$0.6^{\prime\prime}\rm{pixel}^{-1}$. The telescope was autoguided
during the observations, which were performed with the defocussing
technique \citep{southworth2009al1}.

Another complete transit of WASP-67\,b was observed on 2013 June
22, using the DFOSC imager mounted on the 1.54-m Danish Telescope,
also at the ESO observatory in La Silla, during the 2013 observing
campaign by the MiNDSTEp consortium \citep{dominik2010}. The
instrument has a E2V44-82 CCD camera with field of view of
13.7\am$\times$13.7\am\ and a plate scale of
0.39\as\,pixel$^{-1}$. The observations were performed through a
Bessel $R$ filter, the telescope was defocussed and autoguided,
and the CCD was windowed to reduce the readout time. With the
applied defocus, the diameter of the PSF of the target and
reference stars was $\sim$$12^{\as}$, similar to that for the
GROND images.

The optical data collected from both telescopes were reduced using
{\sc defot}, an {\sc idl}\footnote{{\sc idl} is a trademark of the
ITT Visual Information Solutions: {\tt
http://www.ittvis.com/ProductServices/IDL.aspx}} pipeline for
time-series photometry \citep{southworth2009al1}. The images were
debiased and flat-fielded using standard methods, then subjected
to aperture photometry using the {\sc aper}\footnote{{\sc aper} is
part of the {\sc astrolib} subroutine library distributed by NASA
on {\tt http://idlastro.gsfc.nasa.gov}.} task and an optimal
ensemble of comparison stars. Pointing variations were followed by
cross-correlating each image against a reference image. The shape
of the light curve is very insensitive to the aperture sizes, so
we chose those which yielded the lowest scatter. The relative
weights of the comparison stars were optimised simultaneously with
a detrending of the light curve to remove slow instrumental and
astrophysical trends. This was achieved by fitting a straight line
to the out-of-transit data for the DFOSC data and with a
fourth-order polynomial for the GROND data (to compensate for the
lack of reference stars caused by the smaller field of view).

The final differential-flux light curves are plotted in
Fig.\,\ref{Fig:01} and tabulated in Table\,\ref{tab:02}. In
particular, in the top panel of Fig.\,\ref{Fig:01} the GROND light
curves are reported superimposed in order to highlight the
differences of the light-curve shape and of the transit depth
along the four passbands. Interestingly, and contrary to what is
expected for higher-inclination systems (e.g.\
\citealt{knutson2007}), the transit depth gradually increases
moving from blue to red bands. This phenomenon happens because the
planet only covers the limb of the star (as this is a grazing
eclipse), which is fainter in the blue part of the optical
spectrum than the red one due to the stronger limb darkening. So
we expect to see shallower eclipses in the bluest bands for this
system.

The DFOSC Bessel-$R$ light curve is shown in the bottom panel of
Fig.\,\ref{Fig:01} superimposed with the GROND Sloan-$r^{\prime}$
light curve and with that from \citet{hellier2012} obtained with
the Euler 1.2-m telescope through a Gunn-$r$ filter. This panel
highlights the slight variation of the transit depth between the
DFOSC and GROND light curves; the Euler data are more scattered
and agree with both. Slight differences can be caused by the
different filters used or by unocculted starspots. The latter
hypothesis suggests a variation of the starspot activity of the
WASP-67\,A during a period of two years, which is reasonable for a
5200\,K star.

Similar to some previous cases
(\citealp{nikolov2012,mancini2013b,mancini2014}), the quality of
the GROND NIR data were not good enough to extract usable
photometry. We were only able to obtain a noisy light curve in the
$J$ band that, considering the particular transit geometry of the
WASP-67 system, returned very inaccurate estimates of the
photometric parameters in the light-curve fitting process (see
next section) in comparison with the optical ones.

\section{Light-curve analysis}
\label{sec:3}
Our light curves were modelled using the {\sc
jktebop}\footnote{The source code of \textsc{jktebop} is available
at: {\tt http:// www.astro.keele.ac.uk/jkt/codes/jktebop.html}}
code \citep[see][and references therein]{southworth2012}, which
represents the star and planet as biaxial spheroids for
calculation of the reflection and ellipsoidal effects and as
spheres for calculation of the eclipse shapes. The main parameters
fitted by {\sc jktebop} are the orbital inclination, $i$, the
transit midpoint, $T_0$, and the sum and ratio of the fractional
radii of the star and planet, $r_{\mathrm{A}}+r_{\mathrm{b}}$ and
$k = r_{\mathrm{b}}/r_{\mathrm{A}}$. The fractional radii are
defined as $r_{\mathrm{A}} = R_{\mathrm{A}}/a$ and $r_{\mathrm{b}}
= R_{\mathrm{b}}/a$, where $a$ is the orbital semimajor axis, and
$R_{\mathrm{A}}$ and $R_{\mathrm{b}}$ are the absolute radii of
the star and the planet, respectively.

Each light curve was analysed separately, using a quadratic law to
model the limb darkening (LD) effect. Due to the difficulty of
measuring accurate LD coefficients in TEP systems with impact
parameters $b \geq 0.8$ \citep{muller2013}, the WASP-67\,A LD
coefficients were fixed to their theoretical values
\citep{claret2004}. We also assumed that the planetary orbit is
circular \citep{hellier2012}. We included in the fits the
coefficients of a linear (DFOSC) or fourth (GROND) polynomial
versus time in order to fully account for the uncertainty in the
detrending of the light curves.

We also considered the two light curves obtained with the Euler
1.2-m and Trappist 0.6-m telescopes, which were reported in
\citet{hellier2012}. In order to present a homogeneous analysis,
we refitted these two light curves using {\sc jktebop} in the same
manner as for our own data.

As in previous work
\citep{mancini2013a,mancini2013b,mancini2013c,mancini2014}, we
enlarged the error bars of the light curve points generated by our
reduction pipeline. Such a process is necessary because the {\sc
aper} algorithm, used to perform aperture photometry, tends to
underestimate the true uncertainties in the relative magnitude
measurements. This a typical situation in time-series photometry,
where additional noise sources such as red noise are not accounted
for by standard error-estimation algorithms
\citep[e.g.][]{CarterWinn09apj}. We therefore rescaled the error
bars for each eclipse to give a reduced $\chi^2$ of
$\chi_{\nu}^{2}=1$, and then again using the $\beta$ approach
\citep[e.g.][]{gillon2006,winn2008,gibson2008}.

\subsection{Orbital period determination}
\label{sec:3.1}

\begin{table}
\caption{Times of mid-transit of WASP-67\,b and their residuals
versus a linear orbital ephemeris.}
\label{tab:03} %
\centering     %
\tiny
\begin{tabular}{lrcc}
\hline\hline
Time of minimum    & Cycle & Residual & Reference  \\
BJD(TDB)$-2400000$ & no.   & (d)      &            \\
\hline %
$55833.60357  \pm 0.00032 $ &    2 &  0.000510 & 1    \\
$55833.60237  \pm 0.00033 $ &    2 & -0.000690 & 2    \\
$56082.78067  \pm 0.00034 $ &   56 & -0.000578 & 3    \\
$56082.78126  \pm 0.00016 $ &   56 &  0.000012 & 4    \\
$56082.78135  \pm 0.00019 $ &   56 &  0.000102 & 5    \\
$56082.78145  \pm 0.00019 $ &   56 &  0.000202 & 6    \\
$56465.77729  \pm 0.00016 $ &  139 & -0.000064 & 7    \\
\hline %
\end{tabular}
\tablefoot{References: (1) Euler 1.2-m telescope
\citep{hellier2012}; (2) Trappist 0.6-m telescope
\citep{hellier2012}; (3) GROND $g^{\prime}$-band (this work); (4)
GROND $r^{\prime}$-band (this work); (5) GROND $i^{\prime}$-band
(this work); (6) GROND $z^{\prime}$-band (this work); (7) Danish
1.52-m telescope (this work)}
\end{table}

\begin{figure*}
\centering
\includegraphics[width=16cm]{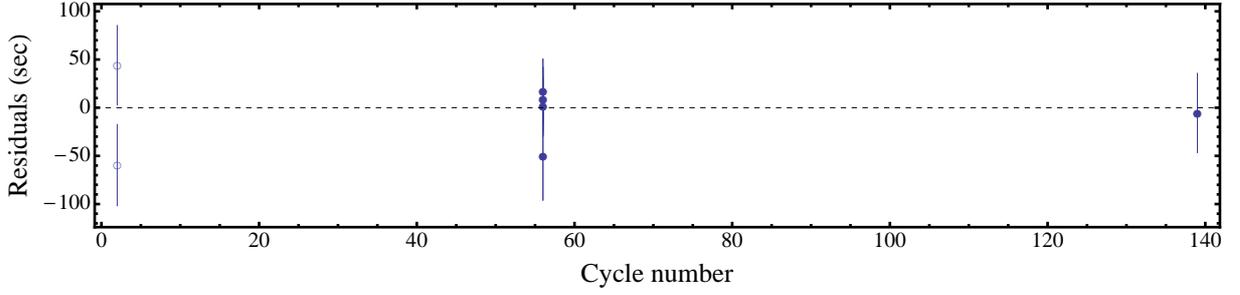}
\caption{Plot of the residuals of the timings of mid-transit of
WASP-67\,b versus a linear ephemeris. The two timings based on the
observations reported by \citet{hellier2012} are plotted using
open circles, while the other timings (this work) are plotted with
filled circles.} \label{Fig:02}
\end{figure*}

We used our photometric data and those coming from the discovery
paper  \citep{hellier2012} to refine the orbital period of
WASP-67\,b. The transit time for each of the datasets was obtained
by fitting with {\sc jktebop}, and uncertainties were estimated
using Monte Carlo simulations. All timings were placed on the
BJD(TDB) time system and are summarised in Table\,\ref{tab:03}.
The plot of the residuals is shown in Fig.\,\ref{Fig:02}. The
resulting measurements of transit midpoints were fitted with a
straight line to obtain a final orbital ephemeris:
\begin{equation}
T_{0} = \mathrm{BJD(TDB)} 2\,455\,824.37424(22) +
4.6144109(27)\,E, \nonumber
\end{equation}
where $E$ is the number of orbital cycles after the reference
epoch, which we take to be that estimated by \citet{hellier2012},
and quantities in brackets denote the uncertainty in the final
digit of the preceding number. The quality of fit,
$\chi_{\nu}^2=1.90$, indicates that a linear ephemeris is not a
perfect match to the observations. However, considering that our
timings cover only three epochs, it is difficult to claim
systematic deviations from the predicted transit times. Future
\emph{Kepler} data will enlarge the number of observed transit
events of WASP-67\,b and may rule in or out possible transit
timing variations.

\subsection{Photometric parameters}
\label{sec:3.2}

\begin{figure*}
\centering
\includegraphics[width=16cm]{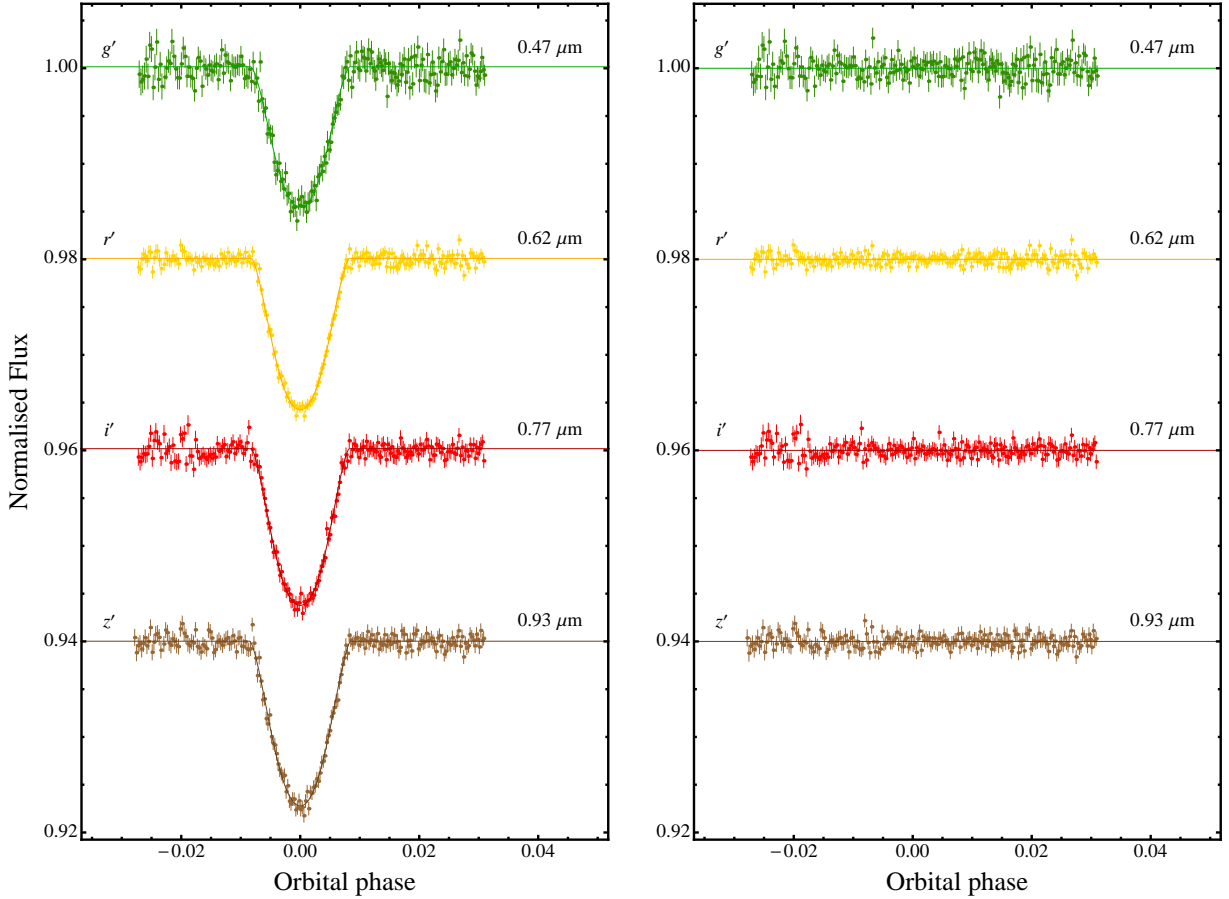}
\caption{\emph{Left-hand panel:} simultaneous optical light curves
of the WASP-67 eclipse observed with GROND. The {\sc jktebop} best
fits are shown as solid lines for each optical data set. The
passbands are labelled on the left of the figure, and their
central wavelengths are given on the right. \emph{Right-hand
panel:} the residuals of each fit.} \label{Fig:03}
\end{figure*}

\begin{figure*}
\centering
\includegraphics[width=16cm]{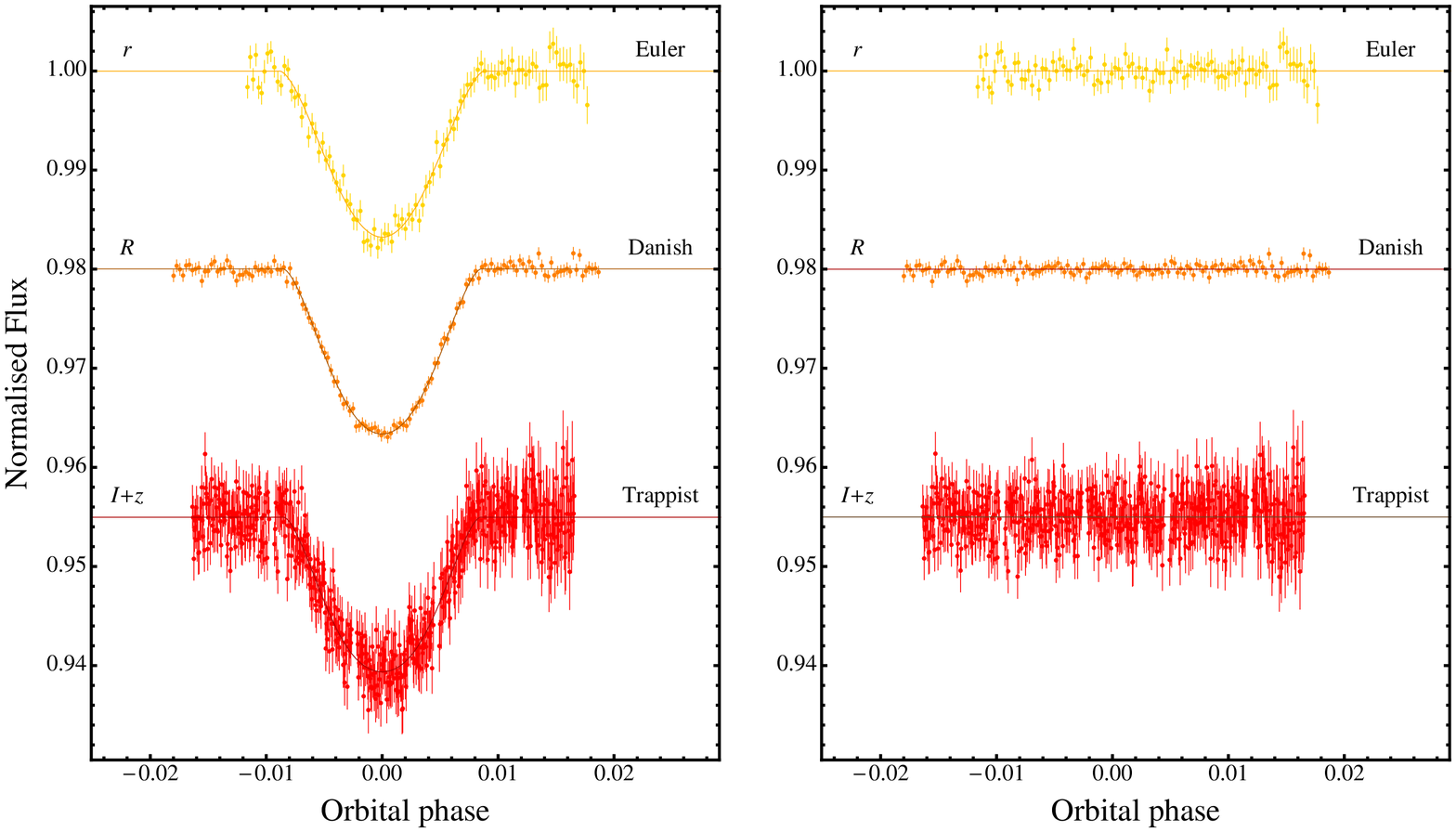}
\caption{\emph{Left-hand panel:} Light curves of the WASP-67
eclipses observed in Gunn-$r$ with the Euler telescope
\citep{hellier2012}, in Bessell-$R$ with the Danish telescope
(this work) and with an $I+z$ filter with the TRAPPIST telescope
\citep{hellier2012}. The filters and the name of each telescope
are labelled on the figure. The {\sc jktebop} best fits are shown
as solid lines for each optical dataset. \emph{Right-hand panel:}
the residuals of each fit.} \label{Fig:04}
\end{figure*}

The GROND light curves and the {\sc jktebop} best-fitting models
are shown in Fig.\,\ref{Fig:03}. A similar plot is reported in
Fig.\,\ref{Fig:04} for the light curves from the Danish Telescope
and \citet{hellier2012}. The parameters of the fits are given in
Table\,\ref{tab:04}. Uncertainties in the fitted parameters from
each solution were calculated from 5500 Monte Carlo simulations
and by a residual-permutation algorithm \citep{southworth08}. The
larger of the two possible error bars was adopted for each case.
The errorbars for the fits to individual light curves are often
strongly asymmetric, due to the morphology of the light curve. The
final photometric parameters were therefore calculated by
multiplying the probability density functions of the different
values. This procedure yielded errorbars which are close to
symmetric for all photometric parameters, which are given in
Table\,\ref{tab:04}. The values obtained by \citet{hellier2012}
are also reported for comparison. Due to their lower quality, we
did not use any of the GROND-NIR light curves to estimate the
final photometric parameters of WASP-67.

\begin{table*}
\setlength{\tabcolsep}{4.5pt} %
\caption{Parameters of the {\sc jktebop} fits to the light curves
of WASP-67.} %
\label{tab:04} %
\centering     %
\tiny          %
\begin{tabular}{llccccc}
\hline\hline
Telescope & Filter & $r_{\mathrm{A}}+r_{\mathrm{b}}$ & $k$ & $i^{\circ}$ & $r_{\mathrm{A}}$ & $r_{\mathrm{b}}$  \\
\hline
MPG\,2.2-m    & Sloan $g^{\prime}$ & \er{0.0831}{0.0061}{0.0036} & \er{0.1323}{0.0192}{0.0058} & \er{86.30}{0.20}{0.39} & \er{0.0734}{0.0040}{0.0029} & \er{0.00972}{0.00200}{0.00074} \\
MPG\,2.2-m    & Sloan $r^{\prime}$ & \er{0.0827}{0.0023}{0.0019} & \er{0.1345}{0.0061}{0.0035} & \er{86.31}{0.11}{0.14} & \er{0.0729}{0.0016}{0.0015} & \er{0.00980}{0.00065}{0.00043} \\
MPG\,2.2-m    & Sloan $i^{\prime}$ & \er{0.0823}{0.0027}{0.0020} & \er{0.1337}{0.0061}{0.0034} & \er{85.34}{0.12}{0.17} & \er{0.0726}{0.0020}{0.0016} & \er{0.00970}{0.00069}{0.00044} \\
MPG\,2.2-m    & Sloan $z^{\prime}$ & \er{0.0865}{0.0040}{0.0027} & \er{0.1424}{0.0139}{0.0065} & \er{86.09}{0.17}{0.25} & \er{0.0757}{0.0025}{0.0019} & \er{0.01078}{0.00143}{0.00075} \\[2pt]
Danish\,1.54-m & Bessel $R$        & \er{0.0868}{0.0038}{0.0026} & \er{0.1445}{0.0151}{0.0070} & \er{86.08}{0.16}{0.25} & \er{0.0758}{0.0023}{0.0018} & \er{0.01095}{0.00148}{0.00077} \\[2pt]
Euler\,1.2-m   & Gunn   $r$        & \er{0.102}{0.013}{0.013}  & \er{0.229}{0.150}{0.080} & \er{85.09}{0.86}{0.87} & \er{0.0828}{0.0021}{0.0047} & \er{0.0189}{0.0127}{0.0074} \\[2pt]
Trappist\,0.6-m  &  $I+z$ filter   & \er{0.0854}{0.0054}{0.0035} & \er{0.1310}{0.0149}{0.0047} & \er{86.16}{0.20}{0.35} & \er{0.0755}{0.0038}{0.0028} & \er{0.00989}{0.00164}{0.00065} \\[2pt]
\hline Final results   &           & $0.0846 \pm 0.0012$ & $0.1379 \pm 0.0030$ & $86.20 \pm 0.07$ & $0.07455 \pm 0.00083$ & $0.01023 \pm 0.00034$ \\
\hline
\citet{hellier2012} & & & $0.1345_{-0.0019}^{+0.0048}$ & $85.8_{-0.4}^{+0.3}$ & & \\
\hline
\end{tabular}
\tablefoot{The final parameters, given in bold, are the weighted
means of the results for the datasets. Results from the discovery
paper are included at the base of the table for comparison. The
Euler and TRAPPIST data sets are from \citet{hellier2012}, while
the others are from this work.}
\end{table*}
%

\section{Physical properties} \label{sec:4}

\begin{table*}
\caption{Final physical properties of the WASP-67 planetary
system, compared with results from \citet{hellier2012}. Two sets of errorbars are given for the results from the current work, the former being statistical and the latter systematic.} %
\label{tab:05} %
\centering
\begin{tabular}{l l r@{\,$\pm$\,}c@{\,$\pm$\,}l c}
\hline \hline
\ & \ & \mcc{\bf This work (final)} & \citet{hellier2012}  \\
\hline
Stellar mass                        & $M_{\rm A}$    (\Msun) & 0.829     & 0.050    & 0.037 & $0.87 \pm 0.04$       \\
Stellar radius                      & $R_{\rm A}$    (\Rsun) & 0.817     & 0.019    & 0.012 & $0.87 \pm 0.04$       \\
Stellar surface gravity             & $\log g_{\rm A}$ (cgs) & 4.533     & 0.014    & 0.007 & $4.50 \pm 0.03$       \\
Stellar density                     & $\rho_{\rm A}$ (\psun) & \mcc{$1.522 \pm 0.049$}      & $1.32 \pm 0.15$       \\[2pt]
Planetary mass                      & $M_{\rm b}$    (\Mjup) & 0.406    & 0.033    & 0.012  & $0.42 \pm 0.04$       \\
Planetary radius                    & $R_{\rm b}$    (\Rjup) & 1.091    & 0.043    & 0.016  & \er{1.4}{0.3}{0.2}    \\
Planetary surface gravity           & $g_{\rm b}$    (\mss)  & \mcc{$8.45 \pm 0.83$}        & \er{5.0}{1.2}{2.3}    \\
Planetary density                   & $\rho_{\rm b}$ (\pjup) & 0.292    & 0.036    & 0.004  & $0.16 \pm 0.08$       \\[2pt]
Planetary equilibrium temperature   & \Teq\              (K) & \mcc{$1003 \pm   20$}        & $1040 \pm 30$         \\
Safronov number                     & \safronov\             & 0.0457   & 0.0037   & 0.0007 &                       \\
Orbital semimajor axis              & $a$               (au) & 0.0510   & 0.0010   & 0.0008 & $0.0517 \pm 0.0008$   \\
Age                                 & Gyr               & \ermcc{ 8.7}{12.7}{ 7.3}{ 5.5}{ 8.6} & \er{2.0}{1.6}{1.0} \\
\hline \end{tabular} %
\end{table*}

Similarly to the \emph{Homogeneous Studies} approach \citep[][and
references therein]{southworth2012}, we used the photometric
parameters estimated in the previous section and the spectroscopic
properties of the parent star (velocity amplitude
$K_{\mathrm{A}}=0.056 \pm 0.004$\kms, effective temperature
$T_{\mathrm{eff}}=5200 \pm 100$\,K and metallicity
$\left[\frac{\mathrm{Fe}}{\mathrm{H}}\right]=-0.07 \pm 0.09$;
\citet{hellier2012}), to revise the physical properties of the
WASP-67 system using the {\sc absdim} code.

We iteratively determined the velocity amplitude of the planet
($K_{\mathrm{b}}$) which yielded the best agreement between the
measured $r_{\mathrm{A}}$ and \Teff, and the values of
$R_{\mathrm{A}}/a$ and \Teff\ predicted by a set of theoretical
stellar models for the calculated stellar mass and
$\left[\frac{\mathrm{Fe}}{\mathrm{H}}\right]$. Statistical errors
were propagated by a perturbation analysis, and the overall best
fit was found by evaluating results for a grid of ages. We
assessed the contribution of systematic errors from theoretical
stellar models by running solutions for five different grids of
models
\citep{Claret04aa,Demarque+04apjs,Pietrinferni+04apj,Vandenberg++06apjs,Dotter+08apjs}.
The final set of physical properties was calculated by taking the
unweighted mean of the five sets of values found from the
different stellar models, and the systematic errors were taken to
be the maximum deviation of a single value from the mean. The
physical parameters of the WASP-67 planetary system are given in
Table\,\ref{tab:05}.

Table\,\ref{tab:05} also shows the values obtained by
\citet{hellier2012} for comparison. We find a smaller radius for
the star, which is attributable to the better constraint on the
stellar density from our high-precision light curves. We also
obtain a significantly smaller planetary radius and hence larger
surface gravity and density. This is due partly to the smaller
stellar radius combined with a comparable measurement of $k$
(Table\,\ref{tab:04}), and partly to an inconsistency between the
$R_{\rm A}$, $R_{\rm b}$ and $k$ values found by
\citet{hellier2012}. The latter issue arises because
\citet{hellier2012} quote the median value of each fitted
parameter from Markov Chain Monte Carlo simulations, rather than
giving the set of parameters corresponding to the single
best-fitting link in the Markov chain (D.\ R.\ Anderson, private
communication).

\section{Variation of the planetary radius with wavelength}
\label{sec:5}

\begin{figure*}
\centering
\includegraphics[width=16cm]{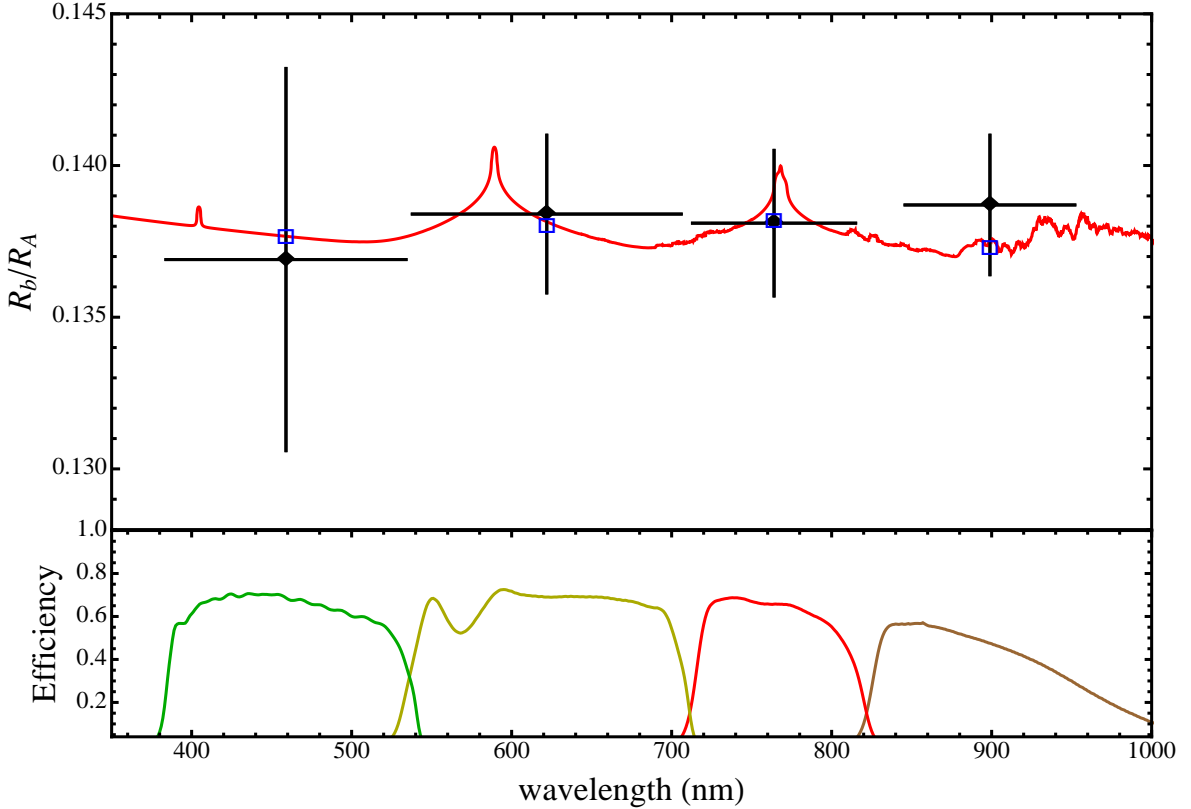}
\caption{Variation of the planetary radius, in terms of
planet/star radius ratio, with wavelength. The black diamonds are
from the transit observations performed with GROND. The vertical
bars represent the errors in the measurements and the horizontal
bars show the FWHM transmission of the passbands used. The
observational points are compared with a synthetic spectrum (see
text for details). Total efficiencies of the GROND filters are
shown in the bottom panel. The blue boxes indicate the predicted
values for the model integrated over the passbands of the
observations.} \label{Fig:05}
\end{figure*}

If it were not for the difficulty of measuring its radius,
WASP-67\,b would be a good target for studies of the planetary
atmosphere due to its low surface gravity. However, its moderate
equilibrium temperature ($T_{\rm eq}^{\,\prime}=1003 \pm 20$\,K)
indicates that the planet should belong to the pL class
\citep{fortney2008}, implying that we do not expect to measure
large variations of the planet radius with wavelength. As the
GROND instrument is able to cover different optical passbands, we
used our data to probe the terminator region of the planetary
atmosphere.

Following our method in previous works
\citep{southworth2012al,mancini2013b}, we refitted the GROND light
curves with all parameters except $k$ fixed to the final values
given in Table\,\ref{tab:04}. This approach maximizes the
precision of estimations of the planet/star radius ratio by
removing common sources of uncertainty. We find the following
values:
$k=0.1369 \pm 0.0063$ for $g^{\prime}$, %
$k=0.1384 \pm 0.0026$ for $r^{\prime}$, %
$k=0.1381 \pm 0.0024$ for $i^{\prime}$, %
$k=0.1387 \pm 0.0023$ for $z^{\prime}$. %
These results are shown in Fig.\,\ref{Fig:05}, where the vertical
errorbars represent the relative errors in the measurements and
the horizontal errorbars show the FWHM transmission of the
passbands used. For illustration, we also show the predictions
from a model atmosphere calculated by \citet{fortney2010} for a
Jupiter-mass planet with a surface gravity of
$g_{\mathrm{b}}=10$\,m\,s$^{-2}$, a base radius of $1.25 \,
R_{\mathrm{Jup}}$ at 10\,bar, and $\Teq = 1000$\,K. The opacity of
strong-absorber molecules, such as gaseous titanium oxide (TiO)
and vanadium oxide (VO), was removed from the model. Our
experimental points are in agreement with the prominent absorption
features of the model (sodium at $\sim$590\,nm and potassium at
$\sim$770\,nm) and, being compatible with a flat transmission
spectrum, do not indicate any large variation of the WASP-67\,b's
radius.

\section{High-resolution image}
\label{sec:High-resolution_image}
Eclipsing binary star systems are a common source of false
positives for transiting planets detected by wide-field
photometry. The host star can have a gravitationally bound
companion, or its light can be contaminated by a background
eclipsing binary which is coincidentally at the same sky position.
Both cases can mimic a planetary-transit signal. Faint nearby
stars may also contaminate the PSF of the target star, thus
slightly lowering the depth of the transit and causing us to
underestimate the radius of both the TEP and its host star.
Finally, these faint nearby stars could also affect the radial
velocity measurements of the star, and thus the measured mass of
the planet (e.g.\ Buchhave et al., 2011).

In order to check if WASP-67\,A is contaminated by any faint
companion or background stars we observed it on 2014/04/21 with
the Andor Technology iXon+ model 897 EMCCD Lucky Camera mounted at
the Danish 1.54-m telescope. The imaging area of this camera is
$512 \times 512$ pixels, and each 16\,$\mu$m pixel projects to
$0^{\prime \prime}.09$ on the sky, giving a $45 \times 45$
arcsec$^2$ field of view. The camera has a special long-pass
filter with a cut-on wavelength of 650\,nm, which corresponds
roughly to a combination of the SDSS $i^{\prime}+z^{\prime}$
filters \citep{skottfelt2013}.

\begin{figure}
\centering
\includegraphics[width=9cm]{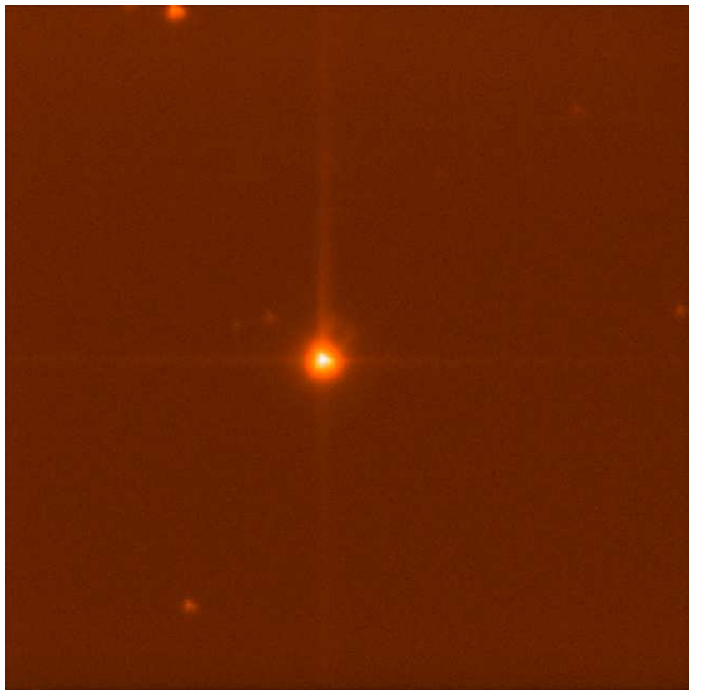}
\caption{Lucky Camera image of WASP-67. The image size is
$45\times45$ arcsec$^2$ and is shown in a logarithmic flux scale
with North up and East to the left. The FWHM of the image is
$0\as.54$. The triangular PSF comes from the telescope in very
good seeing. The extra flux north-west of WASP-67\,A is not a real
contaminating flux source but an optical ghost from the star
caused by internal reflections within the beamsplitters.}
\label{Fig:06}
\end{figure}

\begin{figure}
\centering
\includegraphics[width=9cm]{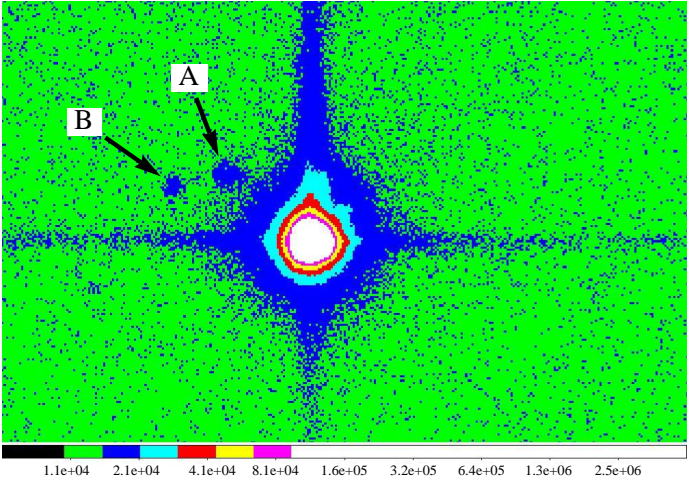}
\caption{Central part of the Lucky camera image in
Fig.\,\ref{Fig:06}. The image is shown in with a logarithmic flux
scale with North up and East to the left. Two faint stars, A at
$\sim4.4^{\prime \prime}$ and B at $\sim6.0^{\prime \prime}$
north-east of WASP-67\,A, are evident. Values in the colour bar
refer to the number of counts in ADU.}\label{Fig:07}
\end{figure}

Fig.\,\ref{Fig:06} shows the resulting image. WASP-67\,A is the
bright star in the centre of the image. Fig.\,\ref{Fig:07} shows
the central region of the image, and it can be seen that two stars
(A and B) occur approximately $4.5\as$ and $6.0\as$ northeast of
WASP-67\,A. The plate scales and inner apertures of DFOSC and
GROND (Table\,\ref{tab:01}) are such that both stars are inside
the defocussed PSFs of WASP-67. However, they are much fainter
than WASP-67\,A, with $\Delta (i^{\prime}+z^{\prime}) = 7.6$\,mag
and 7.9\,mag respectively. They therefore contribute only $0.1\%$
and $0.07\%$ of the total flux in each image, so have a negligible
effect on our results.

In the eventuality that the two faint nearby stars are
intrinsically very blue objects, they could have affected our
$g^{\prime}$-band observations by more than the amount given
above. Measurement of a colour index from multiple high-resolution
images would allow this possibility to be investigated. As a
worst-case scenario, if both contaminants have $\Teff =
30\,000$\,K and are located at such as distance as to contribute
0.1\% of the flux in the Lucky Camera passband, the contamination
in the $g^\prime$-band would be 1.1\%. This figure remains too
small to be important to the current analysis.

\section{Kepler-K2 observations}
\label{K2}

A more extensive study of the WASP-67 planetary system is
anticipated as this object will be observed by the \emph{Kepler}
satellite during its K2 phase. To explore the impact of these
forthcoming observations we have generated a synthetic light curve
matching the K2 data characteristics and subjected it to the same
modelling process as for the real data presented in the current
work.

We calculated a model light curve for the best-fitting photometric
parameters (Table\,\ref{tab:04}) using {\sc jktebop} and for
quadratic LD coefficients appropriate for the $K_p$ passband
\citep{claret2004}. This was extended over the full duration of
the observations for field \#2 (as the schedule for field \#7 is
not yet set), and numerically integrated to the duration of the
short-cadence (58.8\,s) and long-cadence (29.4\,min) data types
obtained by {\it Kepler}. Gaussian random noise was added to each
datapoint equivalent to a scatter 100 parts per million per
six-hour time interval \citep[][their fig.\,10]{howell2014}.
Datapoints outside orbital phases $-0.02$ to $0.02$ were discarded
for computational convenience.

The synthetic light curves were fitted with {\sc jktebop} using
the same treatment as our real datasets for WASP-67, with the
exception that we numerically integrated the model for the
long-cadence simulated data to match its sampling rate
\citep{Me11mn}. We find that the uncertainties in the resulting
photometric parameters are rather similar between the two
cadences, which is due to the relatively smooth brightness
variation through the partial eclipse of WASP-67. They are also
similar to those of our final parameters in Table\,\ref{tab:04},
suggesting that the {\it Kepler} data will not allow a substantial
improvement in the measured physical properties of WASP-67. This
result was unexpected, but can be explained by the larger scatter
of the {\it Kepler} data (0.83\,mmag for short-cadence) versus our
best light curves (see Table\,\ref{tab:01}).

One possibility which is much better suited to K2 observations is
the detection of the rotational period of WASP-67\,A due to
spot-induced brightness modulations. WASP-67\,A is a cool star
(5200\,K) but no spot modulation was detected in the SuperWASP
light curve to a level of roughly 1\,mmag. The data acquired by K2
may allow the rotational period to be estimated, useful for
dynamical and tidal studies.

\section{Summary and conclusions}
\label{sec:summary}

\begin{figure}
\centering
\includegraphics[width=9cm]{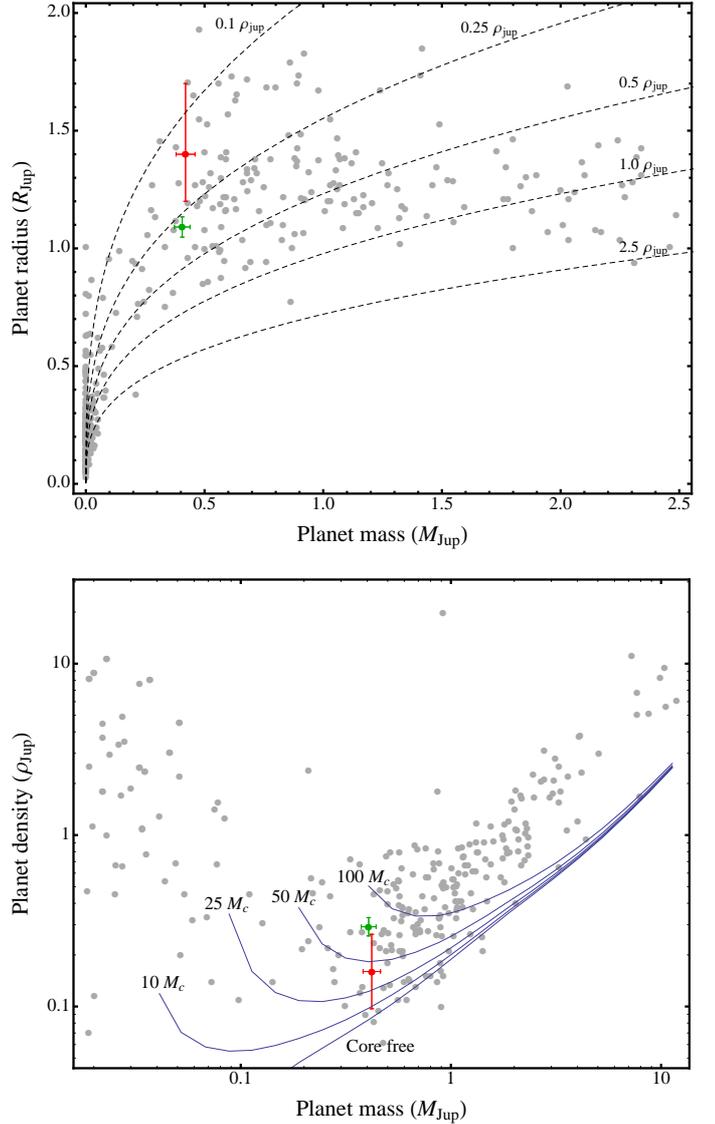}
\caption{\emph{Top panel:} plot of the masses and radii of the
known TEPs. The grey points denote values taken from TEPCat. Their
error bars have been suppressed for clarity. WASP-67\,b is shown
with red \citep{hellier2012} and green (this work) points with
error bars. Dotted lines show where density is 2.5, 1.0, 0.5, 0.25
and 0.1 $\rho_{\mathrm{Jup}}$. \emph{Bottom panel:} the
mass-density diagram of the currently known transiting exoplanets
(taken from TEPCat). Four planetary models with various core
masses and another without a core \citep{fortney2007} are plotted
for comparison.} \label{Fig:08}
\end{figure}

We have presented the first follow-up study of the planetary
system WASP-67, based on the analysis of five new light curves of
two transit events of WASP-67\,b. The first transit was observed
simultaneously with GROND through Sloan $g^\prime$, $r^\prime$,
$i^\prime$, $z^\prime$ filters; the second was observed in
Bessell-$R$ with DFOSC. The transits were monitored roughly one
and two years, respectively, after the reference epoch used by
\citet{hellier2012}. Both transit events were observed in
telescope-defocussing mode, resulting in a photometric precision
of $0.48-1.08$ mmag per observation. We modelled our new and two
published datasets using the {\sc jktebop} code. By estimating the
impact parameter $b$ and the ratio of the planet/star radii, we
found that the criterion for a grazing eclipse, $b+k>1$, is satisfied for all the light curves,
confirming that the eclipse is grazing.

We used the results of the light-curve analysis to substantially
improve the measurements of the physical properties of the planet
and its host star (Table\,\ref{tab:05}). Compared to the discovery
paper \citep{hellier2012}, we find a significantly smaller radius
and a greater density for WASP-67\,b. We obtain $R_{\mathrm{b}} =
1.091 \pm 0.046 \, R_{\mathrm{Jup}}$ versus $1.4_{-0.2}^{+0.3} \,
R_{\mathrm{Jup}}$, and $\rho_{\mathrm{b}} = 0.292 \pm 0.036 \,
\rho_{\mathrm{Jup}}$ versus $0.16 \pm 0.08 \,
\rho_{\mathrm{Jup}}$. Our revised physical properties move
WASP-67\,b into a quite different region of parameter space.
Fig.\,\ref{Fig:08} shows the change in position in the planet
mass-radius plot (top panel) and in the planet mass-density plot
(bottom panel). The revised positions are marked with a green
circle, while the red circle indicates the old values from
\citet{hellier2012}. The values of the other TEPs were taken from
the TEPCat
catalogue\footnote{http://www.astro.keele.ac.uk/jkt/tepcat/.}. For
illustration, the bottom panel of Fig.\,\ref{Fig:08} also shows
10\,Gyr isochrones of exoplanets at 0.045\,AU orbital separation
from a solar analogue \citep{fortney2007}. The plot suggests that
WASP-67\,b should have a more massive core than previously
thought.

As an additional possibility offered by the GROND data, we made an
attempt to investigate possible variations of the radius of
WASP-67\,b in different optical passbands. Our experimental points
are compatible with a flat transmission spectrum and do not
indicate any large variation of the planet's radius. The gradual
increase of the transit depth moving from the GROND $g^{\prime}$
to $z^{\prime}$ band, which is opposite to the case for
higher-inclination systems, is explicable by the fact that
WASP-67\,b only produces grazing eclipses. Due to the stronger
limb darkening, these are shallower in the blue bands than in the
red ones.

\begin{acknowledgements}
This paper is based on observations collected with the MPG 2.2-m
and the Danish 1.54-m telescopes, both located at ESO La Silla,
Chile. Operation of the Danish telescope is based on a grant to
U.G.J.\ by the Danish Natural Science Research Council (FNU).
GROND was built by the high-energy group of MPE in collaboration
with the LSW Tautenburg and ESO, and is operated as a
PI-instrument at the MPG\,2.2-m telescope. We thank the David
Anderson and Coel Hellier for useful discussions, and the referee
for a helpful report. J.S.\ (Keele) acknowledges financial support
from STFC in the form of an Advanced Fellowship. C.S.\ received
funding from the European Union Seventh Framework Programme
(FP7/2007-2013) under grant agreement no.\ 268421. M.R.\
acknowledges support from FONDECYT postdoctoral fellowship
N$^\circ$3120097. T.C.H. would like to acknowledge KASI grant
\#2014-1-400-06. H.K. acknowledges support by the European
Commission under the Marie Curie Intra-European Fellowship
Programme in FP7. S.-H.G.\ and X.-B.W.\ would like to thank the
financial support from National Natural Science Foundation of
China (No.10873031) and Chinese Academy of Sciences (project
KJCX2-YW-T24). O.W.\ thanks the Belgian National Fund for
Scientific Research (FNRS). J.S.\ and O.W.\ acknowledge support
from the Communaut\'{e} francaise de Belgique -- Actionss de
recherche concert\'{e}es -- Acad\'{e}mie universitaire
Wallonie-Europe. K.A., M.D. and M.H. acknowledge grant
NPRP-09-476-1-78 from the Qatar National Research Fund (a member
of Qatar Foundation). This publication was aided by NPRP grant \#
X-019-1-006 from the Qatar National Research Fund (a member of
Qatar Foundation).
We thank the anonymous referee for their useful criticisms and
suggestions that helped us to improve the manuscript.
%
The reduced light curves presented in this work will be made
available at the CDS (http://cdsweb.u-strasbg.fr/). The following
internet-based resources were used in research for this paper: the
ESO Digitized Sky Survey; the NASA Astrophysics Data System; the
SIMBAD data base operated at CDS, Strasbourg, France; and the
arXiv scientific paper preprint service operated by Cornell
University.
\end{acknowledgements}



\begin{thebibliography}{}
%
\bibitem[Anderson et al.(2011)]{anderson2011} %
Anderson, D.~R., Barros, S.~C.~C., Boisse, I., et al. 2011, \pasp,
123, 555
%
\bibitem[B\'{e}ky et al.(2011)]{beky2011} %
B\'{e}ky, B., Bakos, G.~\'{A}., Hartman, J., et al. 2011, \apj,
734, 109
%
\bibitem[Buchhave et al.(2011)]{buchhave2011} %
{Buchhave}, L.~A. , {Latham}, D.~W., {Carter}, J.~A., et al. 2011,
\apjs, 197, 3
%
\bibitem[Carter \& Winn(2009)]{CarterWinn09apj} %
Carter, J.~A., \& Winn, J.~N., 2009, \apj, 704, 51
%
\bibitem[{{Claret}(2004)}]{Claret04aa}
{Claret}, A., 2004, A\&A, 424, 919
%
\bibitem[Claret(2004)]{claret2004} %
Claret, A. 2004, \aap, 428, 1001
%
\bibitem[{{Demarque} et~al.(2004){Demarque}, {Woo}, {Kim}, \&
  {Yi}}]{Demarque+04apjs}
{Demarque}, P., {Woo}, J.-H., {Kim}, Y.-C., {Yi}, S.~K., 2004, ApJS, 155, 667
%
\bibitem[{{Dotter} et~al.(2008){Dotter}, {Chaboyer}, {Jevremovi{\'c}},
  {Kostov}, {Baron}, \& {Ferguson}}]{Dotter+08apjs}
{Dotter}, A., {Chaboyer}, B., {Jevremovi{\'c}}, D., {Kostov}, V., {Baron}, E.,
  {Ferguson}, J.~W., 2008, ApJS, 178, 89
%
\bibitem[Dominik et al.(2010)]{dominik2010}
Dominik, M., J{\o}rgensen, U.~G. , Rattenbury, N.~J., et al. 2010,
AN, 331, 671
%
\bibitem[Fortney et al.(2007)]{fortney2007}
Fortney, J.~J., LMarley, M.~S., Barnes J.~W., 2008, \apj, 678,
1419
%
\bibitem[Fortney et al.(2008)]{fortney2008}
Fortney, J.~J., Lodders, K., Marley, M.~S., Freedman R.~S., 2008,
\apj, 678, 1419
%
\bibitem[Fortney et al.(2010)]{fortney2010}
Fortney, J.~J., Shabram, M., Showman, A.~P., et al. 2010, ApJ,
709, 1396
%
\bibitem[Gillon et al.(2006)]{gillon2006}
Gillon, M., Pont, F., Moutou, C., et al. 2006, \aap, 459, 249
%
\bibitem[Gibson et al.(2008)]{gibson2008}
Gibson, N.~P., Pollacco D., Simpson E.~K., et al. 2008, \aap, 492,
603
%
\bibitem[Greiner et al.(2008)]{greiner2008}
Greiner, J., Bornemann, W., Clemens, C., et al. 2008, \pasp, 120,
405
%
\bibitem[Hellier et al.(2012)]{hellier2012} %
Hellier, C., Anderson, D.~R., Collier Cameron, A., et al. 2012,
\mnras, 426, 739
%
\bibitem[Howell et al.(2014)]{howell2014} %
Howell, S.~B., Sobeck, C., Haas, M., et al. 2014, \pasp, 126, 398
%
\bibitem[Knutson et al.(2007)]{knutson2007} %
Knutson, H.~A., Charbonneau, D., Noyes, R.~W. 2007, \apj, 655, 564
%
\bibitem[Mancini et al.(2013a)]{mancini2013a}
Mancini, L., Southworth, J., Ciceri, S., et al. 2013a, \aap, 551,
A11
%
\bibitem[Mancini et al.(2013b)]{mancini2013b}
Mancini, L., Nikolov, N., Southworth, J., et al. 2013b, \mnras,
430, 2932
%
\bibitem[Mancini et al.(2013c)]{mancini2013c}
Mancini, L., Ciceri, S., Chen, G., et al. 2013c, \mnras, 436, 2
%
\bibitem[Mancini et al.(2014)]{mancini2014}
Mancini, L., Southworth, J., Ciceri, S., et al. 2014, \aap, 562,
A126
%
\bibitem[Nikolov et al.(2012)]{nikolov2012}
Nikolov, N., Henning, Th., Koppenhoefer, J., et al. 2012, \aap,
539, 159
%
\bibitem[M\"{u}ller et al.(2013)]{muller2013}
M\"{u}ller, H.~M., Huber, K.~F., Czesla, S., et al. 2013, \aap,
560, A112
%
\bibitem[{{Pietrinferni} et~al.(2004){Pietrinferni}, {Cassisi}, {Salaris}, \&
  {Castelli}}]{Pietrinferni+04apj}
{Pietrinferni}, A., {Cassisi}, S., {Salaris}, M., {Castelli}, F., 2004, ApJ,
  612, 168
%
\bibitem[Pollacco et al.(2006)]{pollacco2006} %
Pollacco, D. L., Skillen, I., Collier Cameron, A., et al. 2066,
\pasp, 118, 1407
%
\bibitem[Skottfelt et al.(2013)]{skottfelt2013} %
Skottfelt, J., Bramich, D.~M., Figuera Jaimes, R., et al. 2013,
\aap, 553, A111
%
\bibitem[Smalley et al.(2011)]{smalley2011} %
Smalley, B., Anderson, D.~R., Collier Cameron, A., et al. 2011,
\aap, 526, A130
%
\bibitem[Southworth(2008)]{southworth08} %
Southworth, J. 2008, \mnras, 386, 1644
%
\bibitem[Southworth(2011)]{Me11mn}
Southworth, J. 2011, \mnras, 417, 2166
%
\bibitem[Southworth(2012)]{southworth2012}
Southworth, J. 2012, \mnras, 426, 1291
%
\bibitem[Southworth et al.(2009)]{southworth2009al1} 
Southworth, J., Hinse, T.~C., J{\o}rgensen, U.~G., et al. 2009a,
\mnras, 396, 1023
%
\bibitem[Southworth et al.(2012)]{southworth2012al} %
Southworth, J., Mancini, L., Maxted, P.~F.~L., et al. 2012,
\mnras, 422, 3099
%
\bibitem[{{VandenBerg} et~al.(2006){VandenBerg}, {Bergbusch}, \&
  {Dowler}}]{Vandenberg++06apjs}
{VandenBerg}, D.~A., {Bergbusch}, P.~A., {Dowler}, P.~D., 2006, ApJS, 162, 375
%
\bibitem[Winn et al.(2008)]{winn2008}%
Winn, J.~N., Holman, M.~J., Torres, G., et al. 2008, \apj, 683,
1076
%
\bibitem[Winn(2010)]{winn2010} %
Winn J.~N. 2010, in Exoplanet, ed. S. Seager, (The University of
Arizona Press), 56
%
\end{thebibliography}
\end{document}